\documentclass[12pt]{article}
\usepackage{amsmath}
\usepackage{mathtools}
\usepackage{amssymb}
\usepackage{amsfonts}
\usepackage{graphicx}

\title{Optical polarimetry based on geometric phase measurements: unitary Jones matrices}
\date{2019}

\begin{document}
\maketitle
\subsubsection*{Authors:}
Israel Melendez-Montoya, Diana Gonzalez-Hernandez, Alejandra De Luna-Pamanes, and Dorilian Lopez-Mago$^{*}$
\subsubsection*{Addresses:}
Tecnologico de Monterrey, Escuela de Ingenier\'{i}a y Ciencias, Ave. Eugenio Garza Sada 2501, Monterrey, N.L., M\'{e}xico, 64849.\\
$^\ast$ dlopezmago@tec.mx

\section{abstract}
We demonstrate a polarimetry technique based on geometric phase measurements. The technique can be used to obtain either the polarization state of a light beam or the properties of a polarizing optical system. On the one hand, we apply our method to determine the properties of homogeneous and unitary Jones matrices. On the other hand, we provide the recipe to measure the Stokes parameters of a light beam without requiring intensity projections.

\section{Introduction}
The polarization state of a monochromatic light beam can be represented by a point on the Poincar\'{e} sphere. If the light beam propagates through optical elements that change its polarization state, the beam acquires not only a dynamic phase from the optical path length, but also a geometric or Pancharatnam-Berry phase. This geometric phase depends on the trajectory connecting the initial and final points on the Poincar\'{e} sphere \cite{Berry1984,Berry1987}.

The Pancharatnam-Berry phase has applications in the area of structured light \cite{Rubinsztein-Dunlop2017}, which studies the generation of custom optical fields and their interaction with structured materials, such as Pancharatnam-Berry phase optical elements (PBOEs). These PBOEs can be used for wavefront shaping \cite{Marrucci2006}, manipulation of vector beams \cite{Zhou2016,Wang2018} and  waveguides that operate entirely on geometric phases \cite{Slussarenko2015}. Interferometric techniques can be used to measure the geometric phase, as long as the dynamical phase is properly separated\cite{Taylor1992,Loredo2009,Kobayashi2010,VanDijk2010a,Lopez-Mago2017a}. The geometric phase has been experimentally demonstrated for classical light beams and even for quantum states of light\cite{Kwiat1991}.

Optical polarimetry encompasses the theory and techniques to measure the polarization state of light and its interaction with polarizing optical systems.  Jones calculus studies the interaction of polarized light with polarization systems. The Jones matrix of a polarizing optical system provides information about its polarization properties\cite{Savenkov2007}. Its eigenvectors or eigenpolarizations are those polarization states of light that do not change after passing the polarization system. More precisely, the input and output polarization states are equal, but the amplitude and overall phase of the output electric field changes. These changes are characterized by the corresponding eigenvalues or complex transmittances. Jones matrices are classified according to the orthogonality of their eigenpolarizations. If the eigenpolarizations are orthogonal the system is homogeneous, otherwise the system is inhomogeneous\cite{Lu1994}.

Routinely, Stokes and Mueller parameter measurements are performed by several intensity projections\cite{GoldsteinBook, collett2005field}. In this work, we propose and implement a polarimetry technique that relies in measuring the geometric phase instead of intensity projections. Since the geometric phase depends on polarization, we can either determine the polarization state of light or characterize the properties of a polarizing optical system. We focus on polarization systems that can be described by Jones matrices (\textit{i.e.} non-depolarizing elements).

This article is organized as follows. In section \ref{sec:Jones} we explain the method to determine the eigenvectors and eigenvalues of a homogeneous Jones matrix from measurements of the geometric phase. We describe our experiment and numerical analysis to validate our method. In section \ref{sec:Stokes} we develop an alternative application of this idea, which consists on measuring the Stokes parameters of a light beam from geometric phase measurements.

\section{Characterization of homogeneous Jones matrices} \label{sec:Jones}
Consider a polarization system characterized by a homogeneous Jones matrix $\mathbf{J}$. The matrix has two orthonormal eigenpolarizations described by the $2\times 1$  Jones vectors $\mathbf{E}_{1} = [q_{x};q_{y}]$ and $\mathbf{E}_{2}=[-q_{y}^{\ast};q_{x}^{\ast}]$, where $q_{x},q_{y} \in \mathbb{C}$ and $|q_{x}|^{2}+|q_{y}|^{2}=1$. The corresponding normalized Stokes vectors are $\mathbf{Q}_{1}$ and $\mathbf{Q}_{2}$. The orthogonality condition means that $\mathbf{E}_{1}\cdot \mathbf{E}_{2}^{\ast}=0$ and $\mathbf{Q}_{1}=-\mathbf{Q}_{2}=\mathbf{Q}$, with $\mathbf{Q}=[Q_{1};Q_{2};Q_{3}]$. The eigenvalues of $\mathbf{J}$ are $\mu_{1},\mu_{2}\in \mathbb{C}$, therefore $\mathbf{J} \mathbf{E}_{1}=\mu_{1} \mathbf{E}_{1}$ and $\mathbf{J} \mathbf{E}_{2}=\mu_{2} \mathbf{E}_{2}$. If light with polarization $\mathbf{E}_{a}$ passes through $\mathbf{J}$, the state of the resulting field is given by $\mathbf{E}_{b}=\mathbf{J}\mathbf{E}_{a}$. We have demonstrated that the total phase difference between $\mathbf{E}_{a}$ and $\mathbf{E}_{b}$ is equal to\cite{Lopez-Mago2017a,Gutierrez-Vega2011a,Melendez-Montoya2018}
\begin{equation} \label{Eq:totalphase}
\Phi = \mathrm{arg} \left\lbrace \mu_{1} + \mu_{2} + (\mu_{1}-\mu_{2})\mathbf{Q}\cdot \mathbf{A} \right\rbrace,
\end{equation}
where $\mathbf{A}$ is the normalized Stokes vector for $\mathbf{E}_{a}$. Equation (\ref{Eq:totalphase}) may be split into two phases, $\Phi=\Phi_{\mathrm{D}} + \Phi_{\mathrm{G}}$, where $\Phi_{\mathrm{D}}=\mathrm{arg}(\mu_{1}\mu_{2})/2$ is the dynamic phase, and $\Phi_{\mathrm{G}}$  is the geometric phase. 

We notice that we can use Eq. (\ref{Eq:totalphase}) in order to extract the polarization properties of $\mathbf{J}$. In other words, by measuring $\Phi$ as a function of $\mathbf{A}$, we can determine the eigenvectors $\mathbf{Q}_{1},\mathbf{Q}_{2}$ and eigenvalues $\mu_{1},\mu_{2}$ of the polarizing optical system. With that information, the Jones matrix  can be written as\cite{Gutierrez-Vega2011a}
\begin{equation}\label{Eq:JonesMatrix}
\mathbf{J} = \begin{pmatrix}
\mu_{1}|q_{x}|^{2} + \mu_{2}|q_{y}|^{2} & (\mu_{1}-\mu_{2})q_{x}q_{y}^{\ast} \\ (\mu_{1}-\mu_{2})q_{x}^{\ast}q_{y} & \mu_{2}|q_{x}|^{2} + \mu_{1}|q_{y}|^{2}
\end{pmatrix}.
\end{equation}

An important observation of our previous work is that the geometric phase $\Phi_{\mathrm{G}}$ acquired by $\mathbf{A}$ is opposite in sign to the geometric phase $\Phi_{\mathrm{G}}^{\perp}$ acquired by its orthogonal state $\mathbf{A}^{\perp}=-\mathbf{A}$ (only for homogeneous Jones matrices), but the dynamic phase is equal for both\cite{Lopez-Mago2017a}. It means that $\Phi^{\perp} = \Phi_{\mathrm{D}}-\Phi_{\mathrm{G}}$. Therefore,
\begin{equation}\label{Eq:PBphase1}
\Phi-\Phi^{\perp} =2\Phi_{\mathrm{G}} ==  \mathrm{arg} \left\lbrace \mu_{1} + \mu_{2} + (\mu_{1}-\mu_{2})\mathbf{Q}\cdot \mathbf{A} \right\rbrace -  \mathrm{arg} \left\lbrace \mu_{1} + \mu_{2} - (\mu_{1}-\mu_{2})\mathbf{Q}\cdot \mathbf{A} \right\rbrace. 
\end{equation}

For our purposes, we neglect the thickness of the polarization system, which means that the arguments of $\mu_{1}$ and $\mu_{2}$  are purely geometric phases. Furthermore, for homogeneous and non-absorbing polarization systems $|\mu_{1}|=|\mu_{2}|=1$ \cite{Savenkov2007}. Thus, we write $\mu_{1}=\exp(i \delta)$ and $\mu_{2}=\exp(-i\delta)$. Now, Eq. (\ref{Eq:PBphase1}) can be reduced to
\begin{equation} \label{Eq:tanphiPB}
\tan (\Phi_{\mathrm{G}}) =  \mathbf{Q}\cdot \mathbf{A} \tan(\delta).
\end{equation}

The previous equation is the first important result of this work. It gives a simple expression for the geometric phase introduced by \textit{unitary} Jones matrices. In addition, it provides a method to measure either the eigenpolarizations of the optical system or the light polarization state. 

In order to determine the eigenvalues and eigenvectors, we notice that in Eq.(\ref{Eq:tanphiPB}) we have three unknowns: $\delta$ and two components of $\mathbf{Q}$, given that $Q_{1}^{2}+Q_{2}^{2}+Q_{3}^{2}=1$. Therefore, we need at least three measurements of $\Phi_{\mathrm{G}}$ in order to determine the eigenvalues and eigenvectors. 

We do the following procedure to determine $\mathbf{Q}$ and $\delta$ from Eq.(\ref{Eq:tanphiPB}). We perform three measurements with three different polarization states given by the Stokes vectors $\mathbf{A}, \mathbf{B},\mathbf{C}$. Then, we define $\mathbf{\bar{Q}}=\tan(\delta) \mathbf{Q}$. We solve the resulting system of equations
\begin{equation}\label{Eq:matrixtogetQ}
\begin{bmatrix}
A_{1} & A_{2} & A_{3} \\ B_{1} & B_{2} & B_{3} \\ C_{1} & C_{2} & C_{3}
\end{bmatrix} \begin{bmatrix}
\bar{Q}_{1} \\ \bar{Q}_{2} \\ \bar{Q}_{3} 
\end{bmatrix} = \begin{bmatrix}
\tan (\Phi_{\mathrm{G}})_{1} \\ \tan (\Phi_{\mathrm{G}})_{2} \\ \tan (\Phi_{\mathrm{G}})_{3}
\end{bmatrix},
\end{equation}
in order to get $\mathbf{\bar{Q}}$. Then, we obtain $\tan \delta = |\mathbf{\bar{Q}}|=\sqrt{\bar{Q}_{1}^{2}+\bar{Q}_{2}^{2}+\bar{Q}_{3}^{2}}$.

The polarization states that we use to solve Eq. (\ref{Eq:matrixtogetQ}) are the basis polarization states $\mathbf{A} = [1;0;0]$, $\mathbf{B}=[0;1;0]$ and $\mathbf{C}=[0;0;1]$, which are the horizontal, diagonal and circular (right handed) polarization states, respectively.
 
\subsection{Experiment}
\begin{figure}[htbt]
\centering\includegraphics[width=8.4cm]{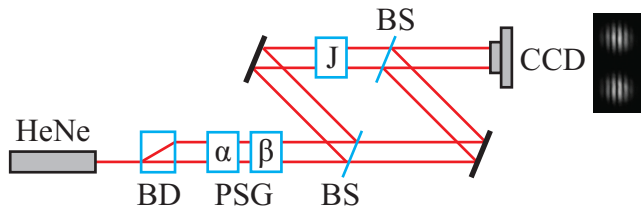}
\caption{(color online) Schematic of our experimental arrangement to measure the geometric phase. Laser source: He-Ne laser. \textbf{BD}: beam displacer. \textbf{PSG}: polarization state generator, which is composed of a quarter-wave plate and a half-wave plate with their fast axes oriented an angle $\alpha$ and $\beta$, respectively.  \textbf{BS}: plate beam splitter. \textbf{CCD}: camera. $\mathbf{J}$: polarizing optical system. The shape of the interferometer is used to minimize the polarization changes from oblique incidence.}\label{Fig:experiment}
\end{figure}

Figure \ref{Fig:experiment} shows our experimental arrangement. The optical setup is composed of three parts: the polarization state generator, a Mach-Zehnder interferometer, and the measurement system\cite{Lopez-Mago2017a}.

The polarization state generator consists of a linearly polarized He-Ne laser, a half-wave plate and a beam displacer followed by a pair of two retarders: a quarter-wave plate $\mathbf{QWP}(\alpha)$ and another half-wave plate $\mathbf{HWP}(\beta)$. The beam displacer generates two parallel beams with horizontal and vertical polarizations and the previous half-wave plate balances their amplitudes. Our beam displacer is a calcite prism (Thorlabs BD40) that provides a beam separation of $4$ mm. We also implemented a tunable beam displacer \cite{Salazar-Serrano2015}, but found some stability issues and decided to keep the calcite prism. The beam separation should be enough to avoid the overlap between the beams. Since our beam diameter is about $1$ mm, the beam displacing prism works fine.  

The plates $\mathbf{QWP}(\alpha)$ and $\mathbf{HWP}(\beta)$ are mounted in a fully automated rotation mount. By properly orienting their fast axes, we transform the input beams into two elliptical and orthogonal polarization states. By using the Jones matrices \cite{collett2005field}
\begin{equation} \label{Eq:Qmatrix}
\mathbf{QWP}(\alpha) = \frac{1}{\sqrt{2}} \begin{bmatrix} 1+i\cos(2\alpha) & i\sin(2\alpha) \\ i\sin(2\alpha) & 1-i\cos(2\alpha)          \end{bmatrix},
\end{equation}
\begin{equation}\label{Hmatrix}
\mathbf{HWP}(\beta) =i\begin{bmatrix}
\cos(2\beta) & \sin(2\beta) \\ \sin(2\beta) & -\cos(2\beta)
\end{bmatrix},
\end{equation}
we can show that light with horizontal polarization passing through $\mathbf{HWP}(\beta)\mathbf{QWP}(\alpha)$ will come out with a polarization state described by the normalized $3 \times 1$ Stokes vector
\begin{equation}\label{Eq:Stokesafterbeamgenerator}
\mathbf{S}^{\mathrm{h}} = [\cos(2\alpha)\cos(4\beta-2\alpha); \cos(2\alpha)\sin(4\beta-2\alpha); - \sin(2\alpha) ].
\end{equation}
Similarly, we can show that light with vertical polarization will come out as $\mathbf{S}^{\mathrm{v}}=- \mathbf{S}^{\mathrm{h}}$. It can be seen from Eq. (\ref{Eq:Stokesafterbeamgenerator}) that $\mathbf{S}^{\mathrm{h}}$ spans all points on the surface of the Poincar\'{e} sphere.  

Both beams are introduced in the Mach-Zehnder interferometer in such a way that each one interferes with itself. Two exact copies of each state are created when passing through the $50:50$ non-polarizing plate beam splitter. One of the arms of the interferometer contains the polarization system $\mathbf{J}$ under study. At the second beam splitter, we have the interference between $\mathbf{E}_{a}$ and $\mathbf{E}_{b}=\mathbf{J}\mathbf{E}_{a}$ for each of the input polarization states. 

The rhomboidal shape of the interferometer is used to reduce the unwanted polarization changes coming from non-normal incidence on the beam splitter\cite{Ericsson2005}. We attribute the polarization changes to phase differences between the vertical (TE) and horizontal (TM) modes at the beam splitter after reflection or transmission\cite{Pezzaniti:94}. In order to determine the incidence angle, we monitor the reflection of a diagonally-polarized beam using a commercial polarimeter (Thorlabs PAX1000VIS). We choose the incidence angle in which the polarization of the reflected beam equals the input beam.

We align the interferometer to obtain transverse interference fringes by adding a tilt to one of the mirrors. The observed interference pattern for one of the input beams has the form $I(x) \propto 1 + |\mathbf{E}_{a}\cdot\mathbf{E}_{b}^{\ast}|\cos(x+\Phi_{\mathrm{D}}+\Phi_{\mathrm{G}})$, where $x$ is the perpendicular direction to the fringes. Correspondingly, the interference pattern of the accompanying beam has the form $I^{\perp}(x) \propto 1 + |\mathbf{E}_{a}\cdot\mathbf{E}_{b}^{\ast}|\cos(x+\Phi_{\mathrm{D}}-\Phi_{\mathrm{G}})$. Therefore, by measuring the relative fringe separation we can measure $\Phi_{\mathrm{G}}$.

We record the interference pattern using a CCD camera (Thorlabs DCU223). We process the raw data to eliminate background noise introduced by the optical elements and the camera. Since the noise features have high frequencies compared to the frequency of the fringes, we eliminate them using a digital low-pass spatial frequency filter on the image of the interference pattern. We accomplish this by applying a Fourier transform to the image and multiplying it by a spatial filter (specifically, a two dimensional Gaussian filter). We then apply an inverse Fourier transform to recover the filtered image\cite{Gonzalez:2003:DIP:993475}. 

For each interference pattern (see Fig. \ref{Fig:experiment}), we take the mean profile of the fringes by averaging a region located at the center of each interferogram. As a result, we get the interferograms $I(x)$ and $I^{\perp}(x)$. The relative shift between $I(x)$ and $I^{\perp}(x)$ is obtained by applying the following Fourier analysis. 

\subsection{Fourier analysis}
We use the following numerical analysis to extract the geometric phase from our measurements. The Fourier transform of $f(x)$ is defined by 
\begin{equation}
\hat{f}(k) = \frac{1}{\sqrt{2\pi}} \int _{-\infty}^{\infty} f(x) e^{-i k x} \mathrm{d} x.
\end{equation}
It is well known that the Fourier transform of $f(x)\cos(k_{0}x)$ is
\begin{equation}\label{Eq:Fourieridentitycosine}
\hat{f}(k) = \frac{1}{2} \left[ \hat{f}(k-k_{0})+\hat{f}(k+k_{0})\right].
\end{equation}

In our measurements, we have an interference pattern of the form $F(x) \approx G(x)\cos(\Phi + k_{0}x)$, where $G(x)$ is a Gaussian envelope. By defining $\Phi = \Phi_{D} + \Phi_{\mathrm{G}}$, we can apply the sum identity for the cosine function to get 
\begin{equation}
F(x) \approx G(x)[ \cos(k_{0}x)\cos(\Phi) - \sin(k_{0}x)\sin(\Phi)]. 
\end{equation}
Applying Eq. (\ref{Eq:Fourieridentitycosine}) to the previous expression we obtain that
\begin{equation}
\hat{f} (k) \approx \frac{1}{2} \cos\Phi \left[ \hat{g}(k - k_{0}) + \hat{g}(k + k_{0})  \right] - \sin \Phi \frac{1}{2i} \left[ \hat{g}(k - k_{0})+\hat{g}(k + k_{0}) \right].
\end{equation}
Arranging common terms
\begin{equation}
\hat{f}(k) \approx \hat{g}(k - k_{0})\frac{1}{2}\left[ \cos\Phi + i \sin\Phi \right] + \hat{g}(k + k_{0})\frac{1}{2}\left[ \cos\Phi - i\sin\Phi \right],  
\end{equation}
which simplifies to
\begin{equation}
\hat{f}(k) = \frac{1}{2} \left[\exp(i\Phi) \hat{g}(k-k_{0}) + \exp(-i\Phi)\hat{g}(k+k_{0}) \right].
\end{equation}

In our case, we are working with two fields of the form
\begin{align}
F_{1}(x) = G(x)\cos(k_{0}x+\Phi_{D}+\Phi_{\mathrm{G}}),
\end{align}
\begin{align}
F_{2}(x) = G(x)\cos(k_{0}x+\Phi_{D}-\Phi_{\mathrm{G}}).
\end{align}
The Fourier transform of each field results in 
\begin{equation}
\hat{f}_{1}(k) = \frac{1}{2} \left[ \exp[i(\Phi_{D}+\Phi_{\mathrm{G}})]\hat{g}(k-k_{0}) + \exp[-i(\Phi_{D}+\Phi_{\mathrm{G}})] \hat{g}(k+k_{0})\right], 
\end{equation}
\begin{equation}
\hat{f}_{2}(k) = \frac{1}{2} \left[ \exp[i(\Phi_{D}-\Phi_{\mathrm{G}})]\hat{g}(k-k_{0}) + \exp[-i(\Phi_{D}-\Phi_{\mathrm{G}})] \hat{g}(k+k_{0})\right]. 
\end{equation}
We multiply both functions to separate $\Phi_{\mathrm{G}}$ from $\Phi_{D}$. The result is 
\begin{align}
4\hat{f}_{1}^{\ast}\hat{f}_{2} & = \exp(-i2\Phi_{\mathrm{G}}) \hat{g}^{\ast}(k - k_{0})\hat{g}(k - k_{0}) + \exp(i2\Phi_{\mathrm{G}}) \hat{g}^{\ast}(k + k_{0})\hat{g}(k + k_{0})\\ 
& + \exp(i2\Phi_{D}) \hat{g}^{\ast}(k - k_{0})\hat{g}(k + k_{0}) + \exp(-i2\Phi_{D}) \hat{g}^{\ast}(k + k_{0})\hat{g}(k - k_{0}).
\end{align}

Notice that $\hat{g}^{\ast}(k + k_{0})\hat{g}(k - k_{0})=0$, due to the fact that width of $\hat{g}$ is less than $|k|=k_{0}$ and so these functions do not overlap. Hence, we can rewrite the expression as
\begin{equation}
\hat{f}_{1}^{\ast}(k) \hat{f}_{2}(k) = \frac{1}{4} \left[ \exp(-i2\Phi_{\mathrm{G}})|\hat{g}(k - k_{0})|^{2} + \exp(i2\Phi_{\mathrm{G}})|\hat{g}(k + k_{0})|^{2} \right].
\end{equation}

Finally, we evaluate the previous equation at $|k|=k_{0}$ and determine the phase of the result, which is equal to $2\Phi_{\mathrm{G}}$. In practice, the function $|\hat{f}_{1}^{\ast}(k) \hat{f}_{2}(k)|$ contains two peaks located at $|k|=k_{0}$. We localize those maxima to determine the value of $k_{0}$. 

\subsection{Results}
We analyze the system $\mathbf{QWP}(\varphi_{1})\mathbf{HWP}(\varphi_{2})\mathbf{QWP}(\varphi_{1})$. Notice that the orientation of the quarter-wave plates is equal. This system is equivalent to a linear retarder where the retardance is controlled by the orientation of the half-wave plate.

The retardance $R$ of a homogeneous polarization system is defined as\cite{Lu1994}
\begin{equation}\label{Eq:retardance}
R = | \arg(\mu_{1})-\arg(\mu_{2})|, \qquad 0\leq R \leq \pi.
\end{equation}
Using Eqs. (\ref{Eq:Qmatrix}) and (\ref{Hmatrix}), we can find that the eigenpolarizations of $\mathbf{QWP}(\varphi_{1})\mathbf{HWP}(\varphi_{2})\mathbf{QWP}(\varphi_{1})$ are
\begin{equation}
\mathbf{E}_{1} = \frac{1}{\sqrt{2}} \begin{bmatrix}
\sqrt{1-\sin 2\varphi_{1}} \\ \frac{\cos 2\varphi_{1}}{\sqrt{1-\sin 2\varphi_{1}}}
\end{bmatrix},
\end{equation}
\begin{equation}
\mathbf{E}_{2} = \frac{1}{\sqrt{2}} \begin{bmatrix}
-\frac{\cos 2\varphi_{1}}{\sqrt{1-\sin 2\varphi_{1}}} \\ \sqrt{1-\sin 2\varphi_{1}}
\end{bmatrix},
\end{equation}
with complex eigenvalues given by
\begin{equation}
\mu_{1} = -\exp(i2\left[\varphi_{1}-\varphi_{2}\right]) \, , \;\; \mu_{2} = -\exp(-i2\left[\varphi_{1}-\varphi_{2}\right]).
\end{equation}
According to Eq. (\ref{Eq:retardance}), the retardance of this system is
\begin{equation}\label{Eq:RforJ}
R=|4\varphi_{1}-4\varphi_{2}|.
\end{equation}
We fix the value of $\varphi_{1} = 0$ and measure the retardance of this system as a function of $\varphi_{2}$. 

\begin{figure}[ht!]
\centering\includegraphics[width=8.4cm]{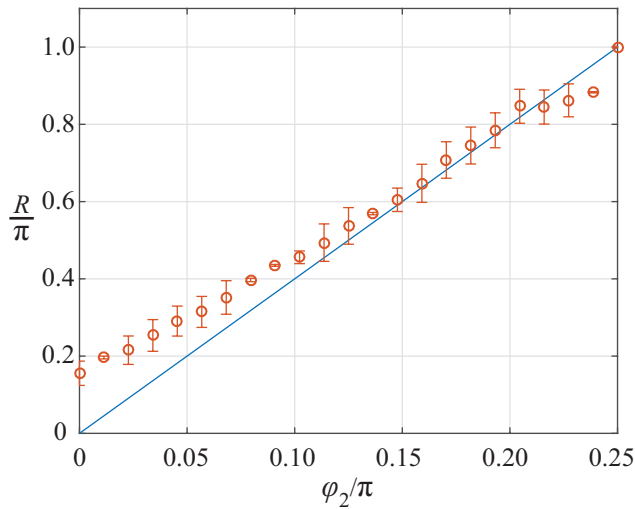}
\caption{(color online) Measurement of the retardance for the homogeneous system $\mathbf{QWP}(\varphi_{1})\mathbf{HWP}(\varphi_{2})\mathbf{QWP}(\varphi_{1})$ as a function of $\varphi_{2}$. The solid line is the theoretical result from Eq. (\ref{Eq:RforJ}) with $\varphi_{1}=0$.}\label{Fig:plotretardance}
\end{figure}

Figure \ref{Fig:plotretardance} shows our experimental results. We applied Eq. (\ref{Eq:matrixtogetQ}) to calculate the eigenpolarizations $\mathbf{Q}$ and $-\mathbf{Q}$ of $\mathbf{J}$, and their corresponding eigenvalues. Our input polarization states are the basis polarizations, \textit{i.e.}, horizontal, diagonal and circular. As seen in Fig. \ref{Fig:plotretardance}, our experimental results do not exactly match the theoretical calculations. We attribute these imperfections to different reasons: (1) the polarization state generator does not produce the exact polarization state due to inaccurate positioning of our rotating mounts, (2) the beams passes through different points on the retarder (close to the border, due to its spatial separation introduced by the beam displacer), alignment of the centroids of the beams, and orientation of the fringe pattern. In addition, when rotating the plates, the beams acquire a slight change on propagation directions, which changes the location of the centroids and introduces a phase error. However, it is observed the linear behaviour of $R$, as predicted by Eq. (\ref{Eq:RforJ}), validating our proposal. 

\section{Measurement of Stokes parameters}\label{sec:Stokes}
Another application of Eq. (\ref{Eq:tanphiPB}) is to measure the Stokes parameters of a light beam. In other words, we determine the polarization state of $\mathbf{A}$ by controlling the values of $\mathbf{Q}$.

We consider three different polarization systems with eigenpolarizations represented by the $3 \times 1$ Stokes vectors $\mathbf{P}$, $\mathbf{Q}$, and $\mathbf{R}$, with eigenvalues $\delta_{P}$, $\delta_{Q}$, and $\delta_{R}$, respectively. Applying Eq. (\ref{Eq:tanphiPB}) and defining $\mathbf{\bar{P}}=\mathbf{P}\tan\delta_{P}$, $\mathbf{\bar{Q}}=\mathbf{Q}\tan\delta_{Q}$, and $\mathbf{\bar{R}}=\mathbf{R}\tan\delta_{R}$, we obtain the equations
\begin{equation}\label{Eq:matrixStokesparameters}
\begin{bmatrix}
\bar{P}_{1} & \bar{P}_{2} & \bar{P}_{3} \\ \bar{Q}_{1} & \bar{Q}_{2} & \bar{Q}_{3} \\ \bar{R}_{1} & \bar{R}_{2} & \bar{R}_{3}
\end{bmatrix} \begin{bmatrix}
A_{1} \\ A_{2} \\ A_{3} 
\end{bmatrix} = \begin{bmatrix}
\tan (\Phi_{\mathrm{G}})_{1} \\ \tan (\Phi_{\mathrm{G}})_{2} \\ \tan (\Phi_{\mathrm{G}})_{3}
\end{bmatrix}.
\end{equation}
By solving the previous equations, we determine the Stokes parameters of the unknown polarization state $\mathbf{A}$. The set of polarization systems that we use to solve Eq. (\ref{Eq:matrixStokesparameters}) are:
\begin{enumerate}
    \item $\mathbf{QWP}(0)$ with eigenvector $[1;0;0]$ and eigenvalue $\exp(i\pi/4)$.
    \item $\mathbf{QWP}(\pi/4)$ with eigenvector $[0;1;0]$ and eigenvalue $\exp(i\pi/4)$.
    \item $\mathbf{HWP}(\pi/8)\mathbf{HWP}(\pi/2)$ with eigenvector $[0;0;1]$ and eigenvalue $\exp(i\pi/4)$.
\end{enumerate}
Notice that all the systems have the same eigenvalues, hence $\delta_{P}=\delta_{Q}=\delta_{R}=\pi/4$. With $\tan(\pi/4)=1$, we find that
\begin{equation}\label{Eq:matrixStokessimple}
\begin{bmatrix}
A_{1} \\ A_{2} \\ A_{3} 
\end{bmatrix} = \begin{bmatrix}
\tan (\Phi_{\mathrm{G}})_{1} \\ \tan (\Phi_{\mathrm{G}})_{2} \\ \tan (\Phi_{\mathrm{G}})_{3}
\end{bmatrix}.
\end{equation}

In order to test our method, we use the same experiment shown in Fig. (\ref{Fig:experiment}). However, we removed the half-wave plate but keep the quarter-wave plate in the polarization state generator. By rotating the quarter-wave plate, the resulting polarization state traces a figure "8" on the Poincar\'{e} sphere\cite{collett2005field}, starting from the horizontal polarization state and passing by both circular polarization states. In this way, we have a representative sample of polarization states to test our method. 

\begin{figure}[ht!]
    \centering
    {\includegraphics[height=8.4cm]{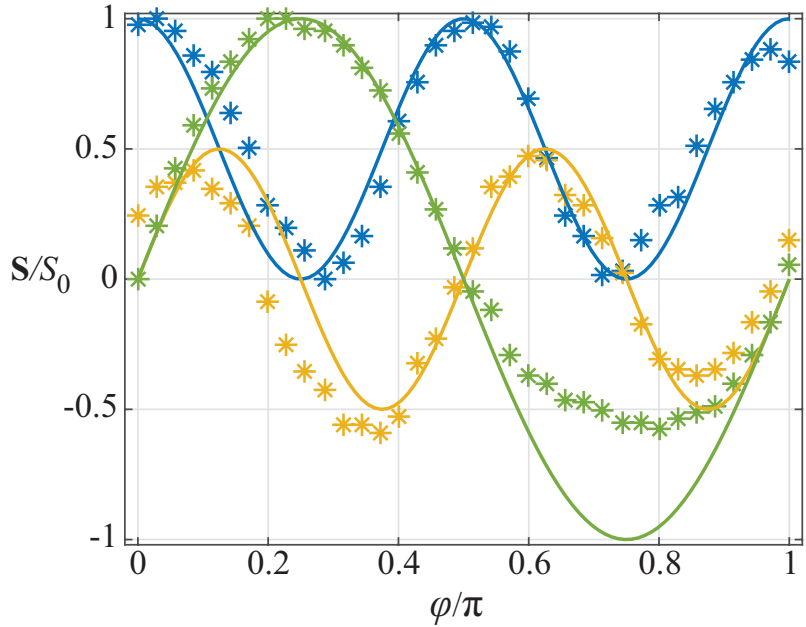} }
    \caption{(color online) Measurement of the normalized Stokes parameters of a light beam after passing a quarter wave plate. We plot each Stokes parameter as a function of the wave plate angle $\varphi$. Solid line is the theoretical curve and the asterisks are the experimental measurements. $s_{1}$: blue, $s_{2}$: yellow, $s_{3}$: green.}
    \label{Fig:stokesparam}
\end{figure}

The procedure to determine an unknown polarization state is the following: (1) to determine the first Stokes parameter $A_{1}$, we put a quarter-wave plate with its fast axis at $0$ degrees and measure the geometric phase, (2) then we rotate the quarter-wave plate at $45$ degrees and measure the geometric phase to determine the second Stokes parameter $A_{2}$, (3) finally, the third Stokes parameter $A_{3}$ is obtained by measuring the geometric phase using two half-wave plates; the first half-wave plate at $90$ degrees and the second half-wave plate at $22.5$ degrees. 

The result of the experiment is shown in Fig. \ref{Fig:stokesparam}, where we plot each Stokes parameter as a function of the quarter-wave plate rotation. For each measurement the geometric phase is obtained using the procedure explained in previous sections, which involves the average of several interferograms and the Fourier analysis. Our results for $s_1$ and $s_2$ agree well with the theory, however, the result for $s_3$ differs after $\pi/2$ and we assert that the reason is due to the quality of our half-wave plates. Nevertheless, the oscillation frequency for each Stokes parameter, as function of $\varphi$, agrees well with the theory.

Notice that the error bars are negligible in this case due to the stability of our system to mechanical vibrations. The measurement of the retardance, shown in Fig. \ref{Fig:plotretardance}, requires three different input polarizations. Therefore, in that case, the error bars come from the imprecision of our polarization state generator. 

\section{Conclusions}
We have shown a polarimetry technique based on measurements of the geometric phase. The applications of this technique are twofold: (i) it can be used to obtain the Jones matrix elements of a polarizing optical system and (ii) it can be used to measure the Stokes parameters of a light beam. Of course, it is questionable if this method, which requires interferometric measurements, is feasible in a practical setting. However, the theory described in Eqs. (\ref{Eq:matrixtogetQ}) and (\ref{Eq:matrixStokesparameters}) is independent of the experiment used to measure the geometric phase, and is valid as long as we can accurately measure $\Phi_{\mathrm{G}}$. Therefore, other techniques to measure the geometric phase can be used, for example, Malhotra \emph{et al.} \cite{Malhotra2018} proposes a method that does not require interferometry. Furthermore, this idea can be applied to other physical systems, equally described by the SU(2) group, \textit{i.e.} any two-level quantum-mechanical system. Inhomogeneous Jones matrices are more involved since the geometric phase between orthogonal polarization states is no longer opposite in sign but completely different, and therefore, our experiment cannot be applied. In addition, the equation for $\Phi_{G}$ becomes a complex transcendental function \cite{Gutierrez-Vega2011a}. Future work is devoted to the characterization of inhomogeneous Jones matrices.

\section*{Funding}
Consejo Nacional de Ciencia y Tecnolog\'{i}a (CONACYT) (Grants: 257517, 280181, 293471).

\bibliographystyle{unsrt}
\bibliography{bibliography}

\begin{thebibliography}{10}

\bibitem{Berry1984}
M.~V. Berry.
\newblock {Quantal Phase Factors Accompanying Adiabatic Changes}.
\newblock {\em Proc. R. Soc. A Math. Phys. Eng. Sci.}, 392:45--57, 1984.

\bibitem{Berry1987}
M.V. Berry.
\newblock {The Adiabatic Phase and Pancharatnam's Phase for Polarized Light}.
\newblock {\em J. Mod. Opt.}, 34:1401--1407, 1987.

\bibitem{Rubinsztein-Dunlop2017}
Halina Rubinsztein-Dunlop, Andrew Forbes, M~V Berry, M~R Dennis, David~L
  Andrews, Masud Mansuripur, Cornelia Denz, Christina Alpmann, Peter Banzer,
  Thomas Bauer, Ebrahim Karimi, Lorenzo Marrucci, Miles Padgett, Monika
  Ritsch-Marte, Natalia~M Litchinitser, Nicholas~P Bigelow,
  C~Rosales-Guzm{\'{a}}n, A~Belmonte, J~P Torres, Tyler~W Neely, Mark Baker,
  Reuven Gordon, Alexander~B Stilgoe, Jacquiline Romero, Andrew~G White, Robert
  Fickler, Alan~E Willner, Guodong Xie, Benjamin McMorran, and Andrew~M Weiner.
\newblock {Roadmap on structured light}.
\newblock {\em Journal of Optics}, 19:013001, 2017.

\bibitem{Marrucci2006}
L.~Marrucci, C.~Manzo, and D.~Paparo.
\newblock {Pancharatnam-Berry phase optical elements for wave front shaping in
  the visible domain: Switchable helical mode generation}.
\newblock {\em Appl. Phys. Lett.}, 88:221102, 2006.

\bibitem{Zhou2016}
Junxiao Zhou, Wenshuai Zhang, Yachao Liu, Yougang Ke, Yuanyuan Liu, Hailu Luo,
  and Shuangchun Wen.
\newblock {Spin-dependent manipulating of vector beams by tailoring
  polarization}.
\newblock {\em Sci. Rep.}, 6:34276, 2016.

\bibitem{Wang2018}
Ruisi Wang, Shanshan He, Shizhen Chen, Jin Zhang, Weixing Shu, Hailu Luo, and
  Shuangchun Wen.
\newblock {Electrically driven generation of arbitrary vector vortex beams on
  the hybrid-order Poincar{\'{e}} sphere}.
\newblock {\em Optics Letters}, 43:3570, 2018.

\bibitem{Slussarenko2015}
Sergei Slussarenko, Alessandro Alberucci, Chandroth~P. Jisha, Bruno Piccirillo,
  Enrico Santamato, Gaetano Assanto, and Lorenzo Marrucci.
\newblock {Guiding light via geometric phases}.
\newblock {\em Nat. Photonics}, 10:571--575, 2016.

\bibitem{Taylor1992}
P~Hariharan and Maitreyee Roy.
\newblock {A Geometric-phase Interferometer}.
\newblock {\em J. Mod. Opt.}, 39:1811--1815, 1992.

\bibitem{Loredo2009}
J.~Loredo, O.~Ort{\'{i}}z, R.~Weing{\"{a}}rtner, and F.~{De Zela}.
\newblock {Measurement of Pancharatnam's phase by robust interferometric and
  polarimetric methods}.
\newblock {\em Phys. Rev. A}, 80:1--9, 2009.

\bibitem{Kobayashi2010}
H.~Kobayashi, S.~Tamate, T.~Nakanishi, K.~Sugiyama, and M.~Kitano.
\newblock {Direct observation of geometric phases using a three-pinhole
  interferometer}.
\newblock {\em Phys. Rev. A}, 81:3--6, 2010.

\bibitem{VanDijk2010a}
Thomas van Dijk, Hugo~F Schouten, Wim Ubachs, and Taco~D Visser.
\newblock {The Pancharatnam-Berry phase for non-cyclic polarization changes.}
\newblock {\em Opt. Express}, 18:10796--804, 2010.

\bibitem{Lopez-Mago2017a}
Dorilian Lopez-Mago, Arturo Canales-Benavides, Raul~I. Hernandez-Aranda, and
  Julio~C. Guti{\'{e}}rrez-Vega.
\newblock {Geometric phase morphology of Jones matrices}.
\newblock {\em Opt. Lett.}, 42:2667, 2017.

\bibitem{Kwiat1991}
Paul~G. Kwiat and Raymond~Y. Chiao.
\newblock {Observation of a nonclassical Berrys phase for the photon}.
\newblock {\em Phys. Rev. Lett.}, 66:588--591, 1991.

\bibitem{Savenkov2007}
Sergey~N Savenkov, Oleksiy~I Sydoruk, and Ranjan~S Muttiah.
\newblock {Eigenanalysis of dichroic, birefringent, and degenerate polarization
  elements: a Jones-calculus study}.
\newblock {\em Appl. Opt.}, 46:6700, 2007.

\bibitem{Lu1994}
Shih-Yau Lu and Russell~A. Chipman.
\newblock {Homogeneous and inhomogeneous Jones matrices}.
\newblock {\em J. Opt. Soc. Am. A}, 11:766, 1994.

\bibitem{GoldsteinBook}
Dennis~H. Goldstein.
\newblock {\em Polarized Light}.
\newblock CRC Press, 2010.

\bibitem{collett2005field}
E.~Collett.
\newblock {\em Field Guide to Polarization}.
\newblock Field Guide Series. Society of Photo Optical, 2005.

\bibitem{Gutierrez-Vega2011a}
Julio~C Guti{\'{e}}rrez-Vega.
\newblock {Pancharatnam-Berry phase of optical systems.}
\newblock {\em Opt. Lett.}, 36:1143--5, 2011.

\bibitem{Melendez-Montoya2018}
Israel Melendez-Montoya, Diana~Laura Gonzalez-Hernandez, Alejandra
  De-Luna-Pamanes, and Dorilian Lopez-Mago.
\newblock Stokes and jones matrix polarimetry based on geometric phase
  measurements.
\newblock {\em Proc. SPIE}, 10749:107491A, 2018.

\bibitem{Salazar-Serrano2015}
Luis~Jos{\'{e}} Salazar-Serrano, Alejandra Valencia, and Juan~P. Torres.
\newblock {Tunable beam displacer}.
\newblock {\em Review of Scientific Instruments}, 86:033109, 2015.

\bibitem{Ericsson2005}
Marie Ericsson, Daryl Achilles, Julio Barreiro, David Branning, Nicholas
  Peters, and Paul Kwiat.
\newblock {Measurement of Geometric Phase for Mixed States Using Single Photon
  Interferometry}.
\newblock {\em Phys. Rev. Lett.}, 94:1--4, 2005.

\bibitem{Pezzaniti:94}
J.~Larry Pezzaniti and Russell~A. Chipman.
\newblock Angular dependence of polarizing beam-splitter cubes.
\newblock {\em Appl. Opt.}, 33:1916--1929, 1994.

\bibitem{Gonzalez:2003:DIP:993475}
Rafael~C. Gonzalez, Richard~E. Woods, and Steven~L. Eddins.
\newblock {\em Digital Image Processing Using MATLAB}.
\newblock Prentice-Hall, Inc., 2003.

\bibitem{Malhotra2018}
T.~Malhotra, R.~Guti\'{e}rrez-Cuevas, J.~Hassett, M.~R. Dennis, A.~N.
  Vamivakas, and M.~A. Alonso.
\newblock {Measuring Geometric Phase without Interferometry}.
\newblock {\em Physical Review Letters}, 120:1--5, 2018.

\end{thebibliography}

\end{document}